\documentclass[aps,amsmath,showpacs,twocolumn]{revtex4}

\usepackage{graphics,epsfig}

\begin{document}


\title{Nucleon-nucleon scattering matrix and its $N_c$ scaling}

\author{Thomas D. Cohen}
\email{cohen@physics.umd.edu}

\author{Vojt\v{e}ch Krej\v{c}i\v{r}\'{i}k}
\email{vkrejcir@umd.edu}

\affiliation{Maryland Center for Fundamental Physics, Department of Physics, \\
 University of Maryland, College Park, MD 20742-4111}

\begin{abstract}
This paper focuses on the scaling of the  S-matrix for elastic nucleon-nucleon scattering at large $N_c$.
It is argued  that the logarithm of a typical S-matrix element  is proportional to $N_c$ in the regime where the large $N_c$ limit is taken with momentum of the incident nucleons proportional $N_c$.   The suggested  scaling was previously derived within the framework of potential scattering models for purely elastic scattering in which the parameters scale the same way as in large $N_c$ QCD.  A variety of heuristic arguments strongly suggest that the same scaling holds in the realistic case where both elastic and inelastic scattering are possible. 

\end{abstract}


\maketitle

\section{Introduction}

Quantum chromodynamics (QCD) is strongly coupled at low and medium energies. This means that the properties of hadrons
 cannot be straightforwardly described within the standard framework of perturbative expansion in the coupling constant.  
At present, it is possible to  compute some hadronic properties directly from QCD using numerical methods on the lattice \cite{lattecereview}, and it will become increasingly possible to compute others with increasing computer power.

However, it is still useful to have analytical tools which give us insights into the underlying dynamics.  
One of the alternative approaches is to work in the limit of a large number of colors and identify the
corrections using the $1/N_c$ expansion.
This approach was originally suggested by t'Hooft \cite{tHooftNC} who realized that the number of colors
may be treated as a free parameter and that in the limiting case of $N_c \rightarrow \infty$ many aspects
of the QCD simplify due to the combinatoric properties of diagrams. The extension
of the large $N_c$ method to baryons was introduced by Witten \cite{WittenNC}.

Although the large $N_c$ limit proved to be quite useful in understanding of qualitative---and in some cases semi-quantitative---aspects of the  meson and baryon physics, its utility for the phenomenology
of nuclear physics is far less clear \cite{GCEM}.
The difficulty is the following: nuclear physics has scales which are much smaller than typical hadronic scales.  So far we know that these scales result from delicate cancellations which occur at $N_c=3$, but which are not expected to hold in general.  A classic example is the  deuteron---the only bound two-nucleon state.  Its binding energy is only 2.2 MeV.  Yet, by $N_c$ counting rules one expects the deepest bound state to be bound by an energy of order $N_c$, {\it i.e.} of a similar scale to the nucleon mass.  

Regardless of its phenomenological utility, it is interesting to consider nuclear phenomena at large $N_c$ since it may give insights into the nuclear problem in a regime where one has a systematic control parameter.  In many ways, the most basic ingredient in nuclear physics is the interaction between nucleons and it is useful to understand such interactions at large $N_c$.  However this raises an interesting question: how does one characterize the interaction in a form which has a well-defined scaling with $N_c$?

To date there have been several ways proposed to characterize these interactions.   The most straightforward way is via the nucleon-nucleon potential \cite{KaplanSavage, KaplanManohar}.   The overall strength of the interaction is large, it scales with $N_c$.  However, the potential has various spin-flavor structures and relative strength of these are fixed by the emergent contracted $SU(4)$ spin-flavor symmetry of baryons at large $N_c$ \cite{GervaisSakita84-1, GervaisSakita84-2, DashenManohar93-1, DashenManohar93-2, DashenJenkinsManohar94, DashenJenkinsManohar95}.  The patterns of strengths seen in phenomenological nucleon-nucleon potential match these rather well.  These patterns are precisely what one would expect from one-meson exchange potentials with appropriate symmetry structure \cite{BanerjeeCohen...}.  
It is interesting to ask whether a meson-exchange picture of nuclear interactions is consistent with large $N_c$ dynamics.   It has been shown that cancellations occur between the crossed diagram  and retardation effects in the box graphs on the level of two-meson exchange; together with the contracted $SU(4)$ symmetry, these cancellations ensures that the desired large $N_c$ behavior is preserved at the order of two-meson exchange \cite{BanerjeeCohen...}.  The situation for higher meson exchange becomes quite complicated, but it appears that the meson-exchange picture is consistent with the $N_c$ scaling of the potentials \cite{Cohen3MesEx1, Cohen3MesEx2}.  

However, there is an important drawback to the use of potentials as a way to characterize the $N_c$ scaling behavior of nucleon-nucleon interactions.  Potentials are not physical quantities by themselves.  They are only inputs into a particular formalism for computing observables and unitarily equivalent theories can have potentials with completely different $N_c$ scaling \cite{Cohen3MesEx2}.  Clearly, it is important to determine the $N_c$ scaling of physical quantities rather than of a theoretical construct---as is the case of the nucleon-nucleon potential.  
The analysis of behavior of physical observables in nucleon-nucleon scattering appears to be a sensible way to do this.

The first description of nucleon-nucleon scattering at large $N_c$ comes from a classic paper of Witten \cite{WittenNC}.  He proposed the use of time-dependent mean-field theory (TDMFT) as the proper tool to study nucleon-nucleon scattering.   In doing so he made a key observation, namely that a smooth large $N_c$ limit does not emerge in TDMFT if one takes the incident momenta of the nucleons to be independent of $N_c$ as in meson-meson scattering.  Rather, to get a smooth large $N_c$ limit one needs 
\begin{equation} \label{WittenKinematics}
p_{\rm incident} \sim N_c \,,
\end{equation}
which, given the $N_c$ scaling of the mass, corresponds to fixed {\it velocity}.  

 TDMFT can be used to deduce the $N_c$ scaling of physical observables. To get a sense of what can be deduced, consider a TDMFT for a model which embodies the $N_c$ scaling---the Skyrme model \cite{Skyrme1}.  TDMFT for the Skyrme model 
 amounts to calculations in classical field theory.  In such a model one needs to set up initial conditions of two solitons approaching each other with fixed impact parameter.   As the two nucleons collide in such calculations, classical pion fields are radiated which corresponds to the emission of pions.   By integrating over the impact parameter and initial spin-flavor orientations, one can deduce quantities such as the outward energy flow as a function of angle (which scales as $N_c^1$) or the baryon number flow 
(which scales as $N_c^0$) \cite{BanerjeeCohen...}.  Since the baryon number is ultimately contained in single baryons and not in the mesons, one can describe the baryon flow in terms of an inclusive differential cross section \cite{CohenGelman02}---namely the differential cross section for the baryon to end up in a certain direction averaging over all possible pion emission.  One interesting result of such an analysis is that the $N_c$ scaling depends on the spin and isospin of the incident nucleons \cite{BanerjeeCohen..., CohenGelman02,  CohenGelman12} due to the contracted $SU(2 N_f)$ symmetry which emerges in large $N_c$ QCD \cite{DashenManohar93-1}.  

However, TDMFT has a fundamental limitation: it cannot predict S-matrix  elements.  Since the S-matrix is at the core of quantum scattering theory, it is particularly important to make contact with it at large $N_c$.  To understand the difficulty with TDMFT, consider again scattering in the context of a Skyrme model which, as noted above, involves the emission of classical pion fields corresponding to the radiation of physical pions.  However, in the calculation one  only has the classical pion field and thus the connection with the particular quantum channels involving the emission of pions is lost.  Moreover, the strength of the pion field in such models is proportional to $\sqrt{N_c}$, the number of pions emitted is proportional to the field strength squared so the number of pions in a given process grows with $N_c$.  Thus, the time-dependent-mean-field theory calculations  describes an average over very different quantum processes as $N_c$ is varied.

This paper focuses on the $N_c$ dependence of the S-matrix for elastic scattering in Witten kinematics of Eq.~(\ref{WittenKinematics}).  Although the nucleon-nucleon interaction is strong, scaling as $N_c^1$, the  S-matrix  in any given partial wave clearly cannot scale linearly with $N_c$ as this would violate unitarity. Recently, it was argued \cite{CohenPRL2012}  that S-matrix elements in any given partial wave  are non-analytic in $N_c$---that  they are scaling exponentially in $N_c$:
\begin{equation}
\log \left( S^{NN}_{\tilde{J}} \right) _{c^i;c^f } \sim N_c \,\,,
\label{logSmat}
\end{equation}
where  $\tilde{J}=J/N_c$ and $J$ specify angular momentum; $c^i$ and $c^f$ represents the additional quantum number for the initial and final states.  These include the spin projections of each of the two particles and the third component of the isospin for each of the two particles. This scaling is important.  From this form it is possible to derive \cite{CohenPRL2012}  the total nucleon-nucleon cross-section:
\begin{equation}
\sigma^{\rm total} = \frac{2 \pi \, \log^2(N_c)}{m_{\pi}^2} \,,
\label{sigmatotal1}
\end{equation}
with the corrections of relative order $\log(\log (N_c))/\log(N_c)$.

The detailed analysis in Ref.~\cite{CohenPRL2012} in which Eq.~(\ref{logSmat}) was first derived is based on a very simplified model: that of spinless nucleons (as one would have for even $N_c$)  in the nonrelativistic regime with momenta of order $N_c$ and interacting via an elastic potential.  The various quantities in the model are taken to scale as the analogous ones do in large $N_c$ QCD.  Clearly such a derivation is heuristic as applied to QCD.  It was argued briefly in Ref.~\cite{CohenPRL2012} that  the scaling is generic and will continue to hold in large $N_c$ QCD.  For this  to be true the complications introduced by large $N_c$ QCD should not change the fundamental result.  There are several of these complications.  The most pressing of these is the existence of inelastic channels which renders the elastic S-matrix non-unitarity.  Other complications include the spin of the nucleon (for $N_c$ odd) and relativistic kinematics.

In this paper we present several arguments that Eq.~(\ref{logSmat}) does in fact hold generally in large $N_c$ QCD.  All of the arguments are, to one extent or another, heuristic.  Taken together, we believe that they form a compelling argument.   This paper is organized as follows:  In the next several sections, we will consider two-flavor QCD and take $N_c$ to be even and nucleons to be spinless and isoscalar.  This simplifies the discussion since the elastic S-matrix  only depends on angle and a treatment via a  partial wave analysis is simple.  The first of these sections will give very simple heuristic arguments that both the real and imaginary part of the logarithm  of the S-matrix scale as $N_c$.  While these are not completely compelling, they are highly suggestive.    These formulations while falling short of a mathematical proof should be sufficient to make it extremely plausible that the logarithm of the S-matrix in Witten kinematics scales as $N_c$.  Following this, is a section in which it is shown that the inclusion of the nucleon's spin does not alter the results and thus Eq.~(\ref{logSmat}) is justified.  Finally, we conclude with a brief discussion of the results and their implications for phenomenology. 

\section{Heuristic Arguments}

This section gives some simple heuristic arguments as to why the logarithm of the S-matrix in Witten kinematics scales as $N_c$.  As noted above,  we will assume  for simplicity that $N_c$ is even and that nucleons are, accordingly, spinless and isoscalar. 

\subsection{A simple scaling argument}

At the simplest level, let us note that the the S-matrix can be expressed as \cite{LandauLifshitz}
\begin{equation} 
\hat{S} = \lim\limits_{t \rightarrow \infty} \,\,  e^{i \hat{H}_{0} t} \,\, e^{- 2 i \hat{H} t} \,\, e^{i \hat{H}_{0} t} \,\,,
\end{equation}
where $\hat{H}$ is the QCD Hamiltonian which encodes the full interaction between hadrons.  $\hat{H}_0$ is the ``free hadronic Hamiltonian'' which governs the propagation of hadrons that do not interact.  This free Hamiltonian is free at the hadronic level and hence is an extremely complicated object in QCD---at the level of quarks and gluons.  However, for our purposes we do not need to construct it explicitly---all that matters are its scaling properties.  According to standard counting rules \cite{WittenNC}, both full Hamiltonian $\hat{H}$ and non-interacting Hamiltonian $\hat{H}_{0}$ are linearly proportional to $N_c$ when acting on the space of states which correspond at early times to wave packets for two nucleons propagating towards each other with Witten kinematics.
Moreover, the difference between the full Hamiltonian $\hat{H}$ and the non-interacting Hamiltonian $\hat{H}_{0}$, which characterizes the interaction itself, is also of order $N_c$ \cite{WittenNC, KaplanManohar}. It is accordingly useful to rewrite $\hat{H}_{0}$ as $N_c \hat{\tilde{H}}_0 $ and $\hat{H}$ as $N_c \left ( \tilde{\hat{H}}_{0} +  \tilde{\hat{H}}_{I} \right)$ where the tildes indicate that the operators are independent of $N_c$.   One thus   is allowed to rewrite the S-matrix as 
\begin{equation} \label{genform}
\hat{S} = \lim\limits_{t \rightarrow \infty} \,\,  e^{i N_c\hat{\tilde{H}}_{0} t} \,\, e^{- 2 i N_c\left ( \tilde{\hat{H}}_{0} +  \tilde{\hat{H}}_{I} \right) t }\,\, e^{i N_c \hat{\tilde{H}}_{0} t} \,\,.
\end{equation}

From Eq.~(\ref{genform}), the interaction acts to affect  the exponential of the S-matrix.
Note, moreover,  that the characteristic range of the nucleon-nucleon interaction is independent of $N_c$ and that the velocity of nucleons in Witten kinematics is also independent of $N_c$.  Thus during the collision process, the nucleons interact for a characteristic  time which is independent of $N_c$.  Since the effect of the interaction on the S matrix occurs during such a time of order one, and the strength of the interaction is of order $N_c$ one expects an overall effect  of the interaction---which occurs in the exponential of the S-matrix---to be order $N_c$.

Of course,  the preceding argument is blatant hand-waving.  The objects under study are quantum operators and one needs to worry that operator ordering effects could radically change the scaling behavior in the exponential.  At the same time, this argument {\it is} rather suggestive.

\subsection{A toy model\label{toy}}

Consider a model in which only elastic nucleon-nucleon scattering is possible, the nucleons are spinless and the system is spherically symmetric so that the angular momentum is conserved.   
In the center-of-mass, the only relevant degrees of freedom are  the three spatial components of the relative distance between the particles.  Suppose further that an artificial parameter $N_c$ is introduced and the parameters of the theory scale with $N_c$ in the same way as do their counterparts in QCD: the mass and the the interaction strength (taken to be for a local interaction) both scale linearly with $N_c$. For simplicity, assume also that only the leading order behavior in $N_c$ is retained.   It is easy to see that the Hamiltonian for the theory is of the form:
\begin{equation} \label{toyham}
H(\vec{q},\vec{p}) \, =  \, N_c \,  \tilde{h} (\vec{q}, \vec{\tilde{p}}) \: \: {\rm with} \; \; \; \vec{\tilde{p}} \equiv \vec{p}/N_c \; .
\end{equation}
This models differs from QCD in one profound way:  it has no inelastic channels.  However it is useful to analyze this model to get insights into  scaling behavior at large $N_c$.

The model in Eq.~(\ref{toyham})  is quite similar to the model analyzed in detail in Ref.~\cite{CohenPRL2012}; however, it is slightly generalized here in that it is not restricted to nonrelativistic kinematics.   In this subsection, the scaling  with $N_c$ of the S-matrix for this model will be derived using a similar but somewhat more general formalism than was given in \cite{CohenPRL2012}; the formalism used here generalizes more straightforwardly to situations where inelasticity is included.

\begin{figure}
\begin{center}

\includegraphics[width=2.5in]{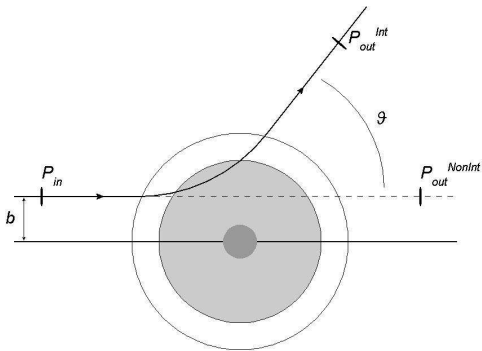}

\caption{A cartoon of the classical scattering.}
\label{classicalscattering}

\end{center}
\end{figure}

First, let us consider the classical equations of motion:
\begin{equation}
\begin{split}
\dot{q_i} &= \frac{\partial \tilde{h}}{\partial \tilde{p}_i} \,\,,\\
\dot{\tilde{p}}_i &= -\frac{\partial \tilde{h}}{\partial q_i} \; .
\end{split}
\end{equation}
The important thing to notice is that the classical equation of motion for $\vec{q}$ and $\vec{\tilde{p}}$ are independent of $N_c$.  Thus, classical scattering trajectories for fixed impact parameter $b$ and fixed initial $\tilde{p}$---such as the one given in Fig.~\ref{classicalscattering}---are independent of $N_c$.   One can calculate the differential cross section for the classical scattering by assuming a random initial distribution of  impact parameters and matching the outgoing scattering angle to the incident impact parameter in the standard way \cite{Goldstein}
\begin{equation} 
 \frac{d \sigma^{\rm classical}}{d \Omega} =  \frac{b(\theta)}{\sin(\theta)} |b'(\theta)| \,\,,
\end{equation}
where $b(\theta)$ is the mapping from the scattering angle to the incident impact parameter.  Thus the classical differential cross section is independent of $N_c$.  

Quantum mechanically, the differential cross section {\it can} depend on $N_c$.   However as $N_c \rightarrow \infty$, the Witten kinematics used here means that the typical momentum in the problem (which is of order $N_c$)  is much larger than the characteristic length scales.  Thus, the large $N_c$ limit {\it is} the semi-classical limit and one expects the classical and quantum differential cross sections to coincide- (except for forward scattering which is outside the semi-classical regime). 
Of course, the differential cross section is determined by the  S-matrix.  However,  the S-matrix contains phase information as well as information about the differential cross-section.  Fortunately, these are related as can be seen by considering Hamilton's action integral along a fixed classical trajectory together with the general quantum mechanical form of the scattering amplitude.  

Recall that for time-independent Hamiltonians  in the semi-classical regime, the quantum phase in a solution of the time-independent 
Schr\"odinger equation associated with a classical trajectory is given (up to additive constants of order unity) by the action.  More precisely, the phase is given by Hamilton's characteristic function $w(\vec{q},\vec{a})$, where $\vec{a}$ is a set of constants of   motion fixed implicitly by the initial conditions.  $w$  is constructed so that $q_i = \partial_{a_i} w(\vec{q},\vec{a})$ and $p_i = \partial_{q_i}  w(\vec{q},\vec{a})$.   The classical equations of motion are fixed by a partial differential equation of the Hamilton-Jacobi form:
\begin{equation}
 H\left (\vec{q}, \nabla_{q}w(\vec{q},\vec{a})  \right ) = E_0 \,\,,
 \end{equation}
 where $E_0$ is the energy.  It is straightforward to see that for trajectories which solve the classical equations of motion, $w$ is given by 
 \begin{equation}
 w_{p,b}(\vec{q}_f)= w_0 + \int_{\vec{q}_0}^{\vec{q}_f }\, \vec{p} \cdot d \vec{q} \; ,
\label{Wint} \end{equation}
 where $\vec{q}_0$ is an arbitrary point chosen to be well before the scattering region, $\vec{q}_f$ is the final point considered and $w_0$ is a constant depending on $\vec{q}_0$ and the constants of motion.  Note that $q_f$ must be along the path of interest, which depends on the angular momentum, $L =  b \, p$.  Quantum mechanically, $L$ corresponds to a partial wave.  The semi-classical phase shift for that partial wave is given by half the difference in phase between two paths--one for the actual path of the particles and the other assuming 
non-interacting particles,  where both paths have the particles  coming in from far away with impact parameter $b$, momentum $p$  and the same initial and final distances from the scattering centers.   Thus, the semi-classical phase shift is given by
 \begin{equation}
\delta_L= \lim_{R \rightarrow \infty} \left(  \frac{1}{2} \int_{\hat{n}^R_0 R}^{\hat{n}^R_f R}\, \vec{p} \cdot d \vec{q}  - \frac{L \, R}{b} \right) \,\,,
 \label{scps}
\end{equation} 
 where $\hat{n}^R_0$ and $\hat{n}^R_f$ are unit vectors in the direction of the initial and final points in the path for any given $R$ with fixed impact parameter $b$.
 Note that each term in Eq.~(\ref{scps}) diverges but the difference is finite.  
 
Now,  consider the $N_c$ scaling behavior of the semi-classical phase-shift.  In Witten kinematics and with a fixed impact parameter, $|\vec{p}| \sim N_c$ for all points along the trajectory and $L \sim N_c$.   Moreover, the region of integration where the interacting and 
non-interacting contributions to the action differ is independent of $N_c$.  Thus  one expects the semi-classical phase shift to be of order $N_c$ in Witten kinematics.  Moreover, Witten kinematics ensures that the system is in the semi-classical regime since the characteristic actions are much greater than unity.  Thus in the large $N_c$ and Witten kinematics with fixed $L/N_c$,  $\delta \sim N_c$.  For the case of a spineless system, this is identical to the scaling in Eq.~(\ref{logSmat}) as desired.   It is important to note that this is fully consistent with the result of Ref.~\cite{CohenPRL2012}.

\subsection{Nonlocality, energy-dependence and the semi-classical limit \label{nonloc} }

How generally does the analysis of the previous section apply?  One important question is whether it should apply for nonlocal potentials.  After all in the context of QCD the nucleon is an extended object with size of order $N_c^0$ and it would not be too surprising if, as a result, the potential generated between nucleons was naturally nonlocal as a result.  However, the scaling result does not generally apply for nonlocal potentials---even ones with a strength  of order $N_c$ and a range of order $N_c^0$.

To understand why, consider a very simple nonrelativistic Hamiltonian for a central local interaction.  In terms of the relative coordinate
\begin{equation}
H=\frac{\vec{p}^2}{M_N} + N_c V(|\vec{x}|) =N_c\left( \frac{\vec{\tilde{p}}^2}{\tilde{M}} +  V(|\vec{x}|) \right ) 
\end{equation}
with $\tilde{M}=M_N/N_c$; the function $V$ is independent of $N_c$.
In Witten kinematics the energy is of order $N_c^1$ so one can write $E = \tilde{E} N_c$.  One can postulate, that the time-independent Schr\"odinger equation $H \psi = E \psi$ has a  solution with a wave function of the form $\psi (\vec{x}) = \langle\vec{x}|\psi\rangle= f (\vec{x}) \exp \left(i N_c w(\vec{x}) \right )$  with $f$ and $w$ independent of $N_c$ and verify its self-consistency.    Note this form ultimately yields  Eq.~(\ref{logSmat}) since it has the phase proportional to $N_c$.  Plugging this form into the the time-independent Schr\"odinger equation and using our simple form of $H$ yields:
\begin{equation}
\begin{split}
&\left \{ N_c^1 \left(  \frac{(\vec{\nabla}w)^2}{\tilde{M}} +  V(|\vec{x}|)            -\tilde{E}      \right ) +  \right .\\
&N_c^0 \left (i \frac{{\nabla}^2 w + 2 \vec{\nabla}w \cdot   \vec{\nabla} \log (f)}{\tilde{M}}      \right ) \\
& \left. N_c^{-1} \left (- \frac{ \left ( \vec{\nabla} \log (f) \right )^2 - {\nabla}^2 \log (f) }{\tilde{M}}      \right ) \right \} \psi=0 \; .
\end{split}
\label{Nexpan}\end{equation}
This {\it is} self consistent.  The leading order part depends on $w$ only.  Thus, insistence  that the order $N_c^1$ contribution vanishes allows for the solution of $w$ given appropriate boundary conditions---one finds it is independent of $N_c$ in both its overall size and its position dependence---as needed for self consistency. Indeed the equation for $w$ {\it is} the classical Hamilton-Jacobi equation. Working at the next orders, $N_c^0$ and $N_c^{-1}$, allows one to use the previously solved $w$ to solve for $f$.   It too can be seen to be independent of $N_c$ in both its overall size and its position dependence.

Now let us look at what happens for the case of a nonlocal potential of the form $N_c V(\vec{x},\vec{x}')$ where the function $V(\vec{x},\vec{x}')$ is independent of $N_c$.  The time-independent  Schr\"odinger equation  is of the form,
\begin{equation}
 -\frac{\nabla^2 \psi(\vec{x}) }{N_c \tilde{M} } + N_c \int {\rm d}^3 x' \, V(\vec{x},\vec{x}') \psi(\vec{x}') = E \psi(\vec{x}) \; .
 \end{equation}
 If one again assumes  $\psi (\vec{x}) =  f (\vec{x}) \exp \left(i N_c w(\vec{x}) \right )$,  one finds that the potential term is of the form 
 \begin{equation}
  N_c \left (\frac{\int {\rm d}^3 x' \, V(\vec{x},\vec{x}')   f (\vec{x'}) \exp \left(i N_c w(\vec{x'}) \right )}{f (\vec{x}) \exp \left(i N_c w(\vec{x}) \right )} \right)  \psi(\vec{x}) \; .
 \end{equation}
 It is clear that the integral will suffer very large cancellations due to the oscillations induced by the complex exponential.  The factor of $N_c$ in the exponent means the rate of oscillation goes to infinity at large $N_c$ leading to a total cancellation at the scale of the leading order.   The actual order of the contribution depends on the detailed form of $V$, but whatever it is, it will be subleading. This is in sharp contrast to the case of a local potential where the potential contribution is of order $N_c$---{\it i.e.} leading order---and leading to phase-shifts which scale as $N_c$.  The effect of interactions for problems with nonlocal potentials of the form $N_c V(\vec{x},\vec{x}')$ with $V(\vec{x},\vec{x}')$ independent of $N_c$ is subleading and the phase shifts will not scale more slowly then $N_c$.  
 
In a certain sense, this should not be too surprising.   The analysis of the model in the previous section depends on the large $N_c$ limit yielding the semi-classical limit.  This is guaranteed to occur at large $N_c$ for Hamiltonians whose leading scaling at large $N_c$ is of the form of Eq.~(\ref{toyham}) for local interactions.  However, non-local interactions do not have a classical analog. Thus it is reasonable that the simple scaling in the classical limit induced at large $N_c$ does not go through for nonlocal interactions.  

However, this does not mean that there cannot be nonlocal interactions yielding the same $N_c$ scaling for phases shifts as in the local case.  For example,  one can imagine making a class of unitary transformation of the theory that preserves the asymptotic wave functions for non-interacting particles (and hence the S-matrix) but  which converts a local Hamiltonian which is of the form of Eq.~(\ref{toyham}) to one which is not.   Thus, the condition for the phase shift to be proportional to $N_c$ is not that the Hamiltonian is necessarily of the form of of Eq.~(\ref{toyham}). 
The sufficient condition is that the Hamiltonian is unitarily equivalent to the one of that form 
(up to corrections which vanish at  large $N_c$).    
 
It is also useful to consider energy-dependent interactions.  These arise naturally when degrees of freedom are integrated out.  Since ultimately a two-nucleon potential   in Witten kinematics must involve integrating out mesonic degrees of freedom, it is important to understand the scaling in this case too.   It is straightforward to see that a theory in which the two-body wavefunction is determined by a local but energy-dependent interaction of the $N_c V(\vec{x};E)$ in a Schr\"odinger equation of the form 
\begin{equation}
\left \{ \frac{- \nabla^2}{M_n} + V(\vec{x};E) \right \} \psi(\vec{x}) = \psi(\vec{x})
\end{equation}
 will have the phase shifts consistent Eq.~(\ref{logSmat}) since the scaling in Eq.~(\ref{Nexpan}) holds self-consistently---just as it does in the energy-independent case.
 
 It is important to reconcile this scaling of the energy dependent potential with the results of Ref.~\cite{Cohen3MesEx2}.  In that work it was shown that, at the level of meson-exchange models, algorithms to extract the nucleon-nucleon potential from higher-order Feynman diagrams  (three-meson exchange and above) did not generically respect the $N_c$ counting for the potential of Ref.~\cite{KaplanSavage, KaplanManohar} but that procedures which systematically remove the energy dependence had the correct scaling.  It is likely, that this came about because these procedures also removed momentum dependence as a by-product.

\subsection{Scaling for the inelasticity}

One deficiency of the simple toy model of  Subsection \ref{toy} is that, by construction, it contains only elastic scattering and thus purely real phase shifts.   In this subsection, we give a general heuristic argument that even when inelasticity is included one also obtains scaling proposed in Eq.~(\ref{logSmat}).  
That is, in the Witten regime with fixed impact parameter (i.e. fixed $ L/N_c$), the imaginary part of the phase shift---like the real part---is of order $N_c$.  Thus, the elastic cross-section in any given partial wave  is exponentially suppressed at large $N_c$.

To see why this should be expected, consider the energy dissipated into the emission of mesons.  It is relatively straightforward to see that in a collision in Witten kinematics, the fraction of the initial kinetic  energy emitted into mesons becomes independent of $N_c$ at large $N_c$---that is the total energy dissipated is of order $N_c^1$ since the initial energy is of order $N_c^1$.  Moreover, the energy distribution of the emitted mesons  also becomes independent of $N_c$ at large $N_c$.  This means that the number of mesons emitted on average, $n_{\rm meson}$, is of order $N_c^1$.    However, ultimately, the process is of quantum mechanical nature and thus probabilistic.   In a   crude sense, one might envision the system making some number of ``attempts'' to create a meson, $n_a $, each with a probability $p$ of creating a meson within each attempt.  Thus, $n_a p= n_{\rm meson} \sim {\cal{O}}(N_c^1)$ and the probability that {\it no} meson is produced---{\it i.e.} that elastic scattering occurs---denoted $P_{\rm elastic}$, is $(1-p)^{n_a}=(1-p)^{n_{\rm meson}/p}$.   
Of course, there are not $n_a$ discrete attempts.  Rather there is a continuous emission of probability.  One can obtain this by taking $n_a \rightarrow \infty$ and $p \rightarrow 0$ with $n_{\rm meson}$ fixed.  This limit yields $P_{\rm elastic}=e^{-{n_{\rm meson}}}$.  Thus one would expect $|S|=\exp(-n_{\rm meson}/2)$ so that $\log{|S|} = -n_{\rm meson}/2 \sim N_c$, which is in agreement with Eq.~(\ref{logSmat}).

It is worth noting that the preceding argument is crude and not correct in detail.  There are two aspects of the argument which are potentially problematic.  The first is that quantum mechanics deals with probability amplitudes rather than probabilities.  The second is that the probabilities considered are associated with the nucleons traveling on a classical trajectory associated with the emission of $n_{\rm meson}$ mesons on average.  Whereas the relevant amplitudes should be associated with the path the nucleons follow when no mesons are emitted.  These paths are different, while the emission of a single meson has a negligible effect on the path, the emission of order $N_c$ mesons makes a change of order unity.  As will be discussed below, these issues do change the value of the probability the reaction is inelastic---but do not alter fundamental scaling of Eq.~(\ref{logSmat}).

Before addressing these issues, it is useful to verify that the number of mesons produced in this kinematic regime  is of order $N_c$.  An easy way to do so is via the Skyrme model \cite{Skyrme1}, which is believed to capture correctly the $N_c$ scaling behavior of QCD.  The Skyrme model is given in terms of a nonlinear sigma model: its dynamical field for two-flavor QCD is a two-by-two unitary matrix $U$.  It is easy to see that in Witten kinematics, classical equations for $U$ with initial conditions for a two-baryon scattering problem are independent of $N_c$.
The solutions involve the classical fields after the collision time having two solitons moving away from each other at some scattering angle and classical radiation of the $U$ field away from the two solitons.  It is straightforward to see that the fraction of energy carried away by such radiation is independent of $N_c$, as advertised, above yielding $n_{\rm meson} \sim N_c$.  
Alternatively, one can see the result from the parametrization of $U$ in terms of the pion field, $U=\exp \left (i \frac{(\vec{\tau} \cdot \pi)}{f_\pi} \right)$. From this definition, it follows that the classical pion fields emitted in such a collision are proportional to $f_\pi$.  Since $f_\pi \sim N_c^{1/2}$, it follows that the pion field strength is also $\sim N_c^{1/2}$. Since the particle number associated with the classical field is proportional to kinematic factors independent of $N_c$ times the field strength squared, it follows that the particle number is proportional to $N_c$.

\subsubsection{Amplitudes vs probability}

The simple argument yielding $|S|=\exp(-n_{\rm meson}/2)$ was based on a statistical treatment of independent ``attempts'' of the system to create mesons.  However these attempts are quantum mechanical amplitudes and not independent probabilities.  The difference between these is associated with quantum mechanical coherence. Thus one might worry that this fact might invalidate the result.   However, this is not the case.

To see why consider a description of the scattering process in QCD in which the fundamental quark and gluon degrees of freedom are replaced by an effective hadronic theory.  While such a theory will be complicated and embody nonlocalities, in principle such a description  should be possible; it will correctly encode the large $N_c$ scaling of the dynamics.  Recall that the large $N_c$ limit is essentially a classical one at the level of hadronic dynamics \cite{WittenNC}.  One might be tempted to take the classical nature of the large $N_c$ dynamics as an indication that the quantum mechanical distinction between amplitudes and probabilities should be unimportant, thereby justifying the argument given above.   However, this is not so obvious.    In fact, to get an accurate description of quantum system in terms of  classical meson fields one needs a very large measure of coherence at the quantum level.   Indeed,  one generally regards a Glauber coherent state \cite{GlauberCoherent} as the ``most classical'' quantum state with fixed expectation value for a field and its conjugate momentum.  Given that we ignored coherence in  the argument given above---focusing on probabilities and not amplitudes---one might worry that the derivation fails when there is  significant coherence.  In fact,  suppose that we describe the final state in two-nucleon scattering in this hadronic language and, to capture the classical nature of the dynamics, model the mesonic sector of the quantum state after the scattering process as a coherent state.  Doing so automatically yields $P_{\rm elastic}=e^{-{n_{\rm meson}}}$ for  the probability that no mesons are present---precisely as found in the purely probabilistic calculation without coherence.

To illustrate this consider, a simplified model which contains only one species of meson which we take to be a scalar meson.  The creation operator for a meson of momentum $\vec{p}$ is denoted $a^\dagger_{\vec{p}}$, and it is normalized to satisfy the commutation relation $[a_{\vec{p}},a^\dagger_{\vec{p}'}] = (2 \pi)^3 \delta^3(\vec{p}-\vec{p}\,')$.   The most general coherent state is of the form 
\begin{equation} 
\begin{split}
|f\rangle & = {\cal N} \exp \left (\int \frac{d^3 p}{(2 \pi)^3}  f(p) a^\dagger_{\vec{p}} \right) |0\rangle  \\ 
  {\cal N} & = \exp \left (-\frac{ \left|\int \frac{d^3 p}{(2 \pi)^3} f(p) \right|^2}{2} \right)
\end{split}
\end{equation}
The number operator $\hat{n}$ is given by $\int \frac{d^3 p}{(2 \pi)^3} a^\dagger_{\vec{p}} a_{\vec{p}} $.  Thus
\begin{equation}
n_{\rm meson} \equiv \langle f|\hat{n}|f \rangle = \int \frac{d^3 p}{(2 \pi)^3} \left | f(p) \right|^2
\end{equation}
On the other hand, the probability that no meson exists in the state is simply $|{\cal N}|^2$ which is given by
\begin{equation}
|{\cal N}|^2= \exp \left (-\left|\int \frac{d^3 p}{(2 \pi)^3} f(p) \right|^2 \right )=\exp \left (- n_{\rm meson} \right ) \; .
\end{equation}
As noted above, this is identical to the expression based on uncorrelated probabilities.  Moreover,  while this result was derived for a model with a single species of meson, it should be obvious that the result would be identical if the coherent state included multiple species of meson.

This result shows that coherence, {\it per se}, does not invalidate the result that the probability of producing no mesons is given by $\exp \left (- n_{\rm meson} \right )$.  However, it does not demonstrate that the result is correct either.  Indeed, the result is {\it not} correct in detail---although the exponential dependence on $N_c$ is.  {\it A priori} there is a reason to suspect that the coherent state is not adequate to describe this component of the wave function of the system.  Recall, that the use of the  coherent state here is as a quantum description of a semi-classical process.  The true quantum description is expected to be much more complicated. However, the coherent state captures the dominant behavior of a system which creates many coherent mesons. The difficulty is that the component of the state in which no mesons are produced is an exponentially small fraction of the coherent state wave function and is thus a highly atypical configuration.  There is no reason for  the coherent state describing the behavior of typical components of the actual quantum state for the semi-classical process to accurately describe highly atypical ones.  Moreover, as will be described below, based on a simple model developed in analogy to bremsstrahlung, there is a very good reason to believe that it does not.

\subsubsection{Meson bremsstrahlung}

The model discussed in this section is motivated by an analogy to electromagnetic bremsstrahlung: a process in which an ion
 undergoing some accelerated motion emits an electromagnetic radiation.
Analogous methods may be used to analyze the emission of mesons from  nucleons when they
are scattered.   In doing this we will assume that the system is in the semi-classical regime.  One can imagine forming an initial wave packet of each of the nucleons involved in the scattering.  If the system is in the Witten limit, the wave packet should not disperse substantially over time scales of the scattering---${\cal O}(N_c^0)$---and hence acts like a classical source for the emission of mesons.  For simplicity of illustration, we will analyze this process in a model with one type of mesons (a scalar) and a spinless nucleon.  The model is designed to capture  the $N_c$ scaling of $QCD$ by taking the meson mass to be of order $N_c^0$, the nucleon mass to be of order $N_c^1$ and the coupling between them (which is taken to be of a Yukawa form) has a strength of order $N_c^{1/2}$.  For simplicity we neglect meson-meson interactions. These simplifications should not alter the fundamental $N_c$ scaling rules.

By assumption the system can be treated semi-classically since it is in the Witten limit.  The semi-classical calculation  has two parts coupled together---the  mesons emission can be treated classically as arising from sources  associated with the acceleration of nucleons   and the nucleon's trajectory can be computed due to forces from the nucleon-nucleon interaction and from a back reaction from the mesons which are emitted.  
Note the  analogy with bremsstrahlung:  the accelerating ion in bremsstrahlung  acts as a classical source for electromagnetic field whereas here an accelerating nucleon can be viewed as
classical source for a meson field.

In studying the presented model, we start by focusing on half of the problem: the emission of mesons taking the nucleons trajectories to be fixed externally.  Ultimately, one needs to choose these trajectories in a self-consistent manner:  these trajectories need to be the ones which emerge taking into account  the back reaction due to the emission of mesons.  However, for the present purposes it is sufficient to assume that the path taken by the nucleons and the rate in which they follow the path are both independent of $N_c$ at large $N_c$ as expected in Witten kinematics \cite{WittenNC}.  The starting point for the analysis is a Lagrangian for a meson field in the presence of a classical source:
\begin{equation}
\mathcal{L}=\frac{1}{2} \left(\partial_{\mu} \phi\right)^2 - \frac{1}{2} m^2 \phi^2 - g  \, J(x) \phi
\end{equation}
where---and it is crucial---the scalar field's coupling to the external current $g$ scales as $N_c^{1/2}$.
The solution of the equation of motion for the meson field is:
\begin{equation}
\phi = g \, \int {\rm d}^4 x' \Delta_R (x-x') J(x'),
\label{EOM}
\end{equation}
where $\Delta_R (x-x')$ is the retarded Green's function and $J(x')$ represents the source localized at the classical trajectory followed by the baryon
in consideration.

On the classical level, the energy flux carried by the meson fields is given by the $(0i)^{\rm th}$ component of the stress-energy tensor
(the analog of the Poynting vector for the electromagnetic field):
\begin{equation}
S_i= \dot{\phi}\, \phi_{,i},
\label{SI}
\end{equation}
and the overall outgoing energy is calculated by integration over a surface surrounding the area of interest.
The exact formula for the outgoing energy flux carried by mesons  depends on the specific trajectory of the
nucleon and is quite complicated. However, the $N_c$ scaling can be deduced straightforwardly.
Expression for the energy flux $S_i$ (\ref{SI}) is quadratic in the field $\phi$ (\ref{EOM}), which is linear in the coupling constant $g$.
Since the trajectories are, by hypothesis, independent of $N_c$,  $g$ contains the only $N_c$ dependence in the problem. From the scaling shown above, it directly follows that the energy flux
is proportional to $N_c$:
\begin{equation}
\int\limits_{\Omega} \vec{S} \sim g^2 \sim N_c \,.
\end{equation}

The discussion above was based on the classical field theory point of view.
Doing the proper quantization of the field in Eq. (\ref{EOM}) is complicated and is beyond the scope of the presented paper.
Fortunately, deducing the $N_c$ scaling is possible and is precisely what was quoted earlier:
the total energy carried by the field is of order $N_c^1$. Since the mass and energy of a meson are
of order $N_c^0$ in Witten kinematics, the number of particles emitted when baryon follows an accelerated classical trajectory
must be of order $N_c^1$.

\begin{figure}
\begin{center}

\includegraphics[width=2in]{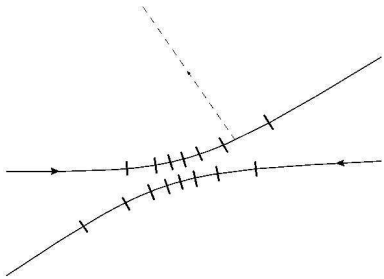}

\caption{A cartoon representing the classical path of the baryons during the scattering and an emission of one meson from one segment of the path}
\label{isoscadiagrams}

\end{center}
\end{figure}

To describe the emission process of individual mesons, let us divide the trajectory for the two baryons into $s$ small segments each corresponding to a certain small time interval in such a way that the probability of emitting one meson $p$ in this segment is small and probability of emitting more than one meson is negligible.  One can choose the size of these segments as such that $p$ is the same in each segment. The possible emission of a meson via bremsstrahlung in one of these segments is a concrete model for the  ``attempts'' discussed earlier. The analysis goes through exactly as before: $n_{\rm meson} = p s \sim N_c$ and  $P_{\rm elastic}=\exp( -n_{\rm meson}) $.

There is a subtlety which must be addressed, however.   The preceding analysis was based on a fixed external trajectory for the baryons.   Let us label the trajectory by $T$ and the expected number of meson associated with that trajectory as $n_{\rm meson}^T$.   Note that  $n_{\rm meson}^T$ describes the number of mesons emitted on average if a source was forced to follow trajectory $T$---not necessarily the number of mesons created on average in the physical process.
The precise value of $n_{\rm meson}^T$ clearly depends on {\it which} trajectory is chosen since the bremsstrahlung process depends on acceleration.  However, it is also clear that provided that $T$ is independent of $N_c$ in both its path and the rate it follows the path, that  $n_{\rm meson}^T \sim N_c$ and thus $\log (P_{\rm elastic}) \sim N_c$.

 The correct choice for describing typical outcomes---those in which the number of mesons produced is close to the average produced by the process $n_{\rm meson}$ where by close we mean differing from it by order $n_{\rm meson}^{1/2}$---is quite clear.  In that case, the correct path to use is the classical one, $T_{\rm classical}$.  This path includes the effects of the force due to the classical back reaction of the mesons on the nucleon.  Thus in the large $N_c$ limit,   $n_{\rm meson}=n_{\rm meson}^{T_{\rm classical}}$.  However, this is clearly {\it not} the appropriate path to consider when describing purely elastic scattering.  In the elastic scattering case, there is no back reaction and the initial and final kinetic energy of the nucleons are equal.  The underlying quantum mechanical point is that the meson degrees of freedom and the baryon degrees of freedom are correlated.  The correlations are such that for the bulk of the components in the state a semi-classical description is valid.  However, if one were to focus on the small part of the state associated with elastic scattering, there is still a semi-classical description one could make for the motion of the nucleons.  To find the classical trajectory, simply do the standard classical calculation but systematically remove the effects of back reaction.  Let us denote this trajectory $T_{\rm nbr}$, (where nbr stands for no back reaction).  Let's define   the average number of mesons which {\it  would have been} produced had the sources been forced to follow $T_{\rm nbr}$ as  $n_{\rm meson}^{T_{\rm nbr}}$.  Thus one sees that $P_{\rm elastic} = \exp( -n_{\rm meson}^{T_{\rm nbr}}) \ne  \exp( -n_{\rm meson})$.  
 
 The result that  $P_{\rm elastic} \ne  \exp( -n_{\rm meson})$ is presumably generically correct.  Nevertheless, note that in Witten kinematics $T_{\rm nbr}$ will be independent of $N_c$ (just as $T_{\rm classical}$ is) which in turn implies that $n_{\rm meson}^{T_{\rm nbr}} \sim N_c$.  This in turn implies  that $\log (P_{\rm elastic}) \sim N_c$ as was true in the simple analysis yielding $P_{\rm elastic} =  \exp( -n_{\rm meson})$.  Thus, although the simple analysis yielding $P_{\rm elastic} =  \exp( -n_{\rm meson})$ is not correct in detail, it does correctly describe the $N_c$ scaling.  
 
 To summarize this line of reasoning, models at the hadronic level which encode the correct large $N_c$ scaling of QCD but do not include spin and flavor degrees of freedom, when treated semi-classically reproduce the scaling in Eq.~(\ref{logSmat}).

\subsubsection{Complex Potentials}

The preceding analysis of inelasticities was somewhat heuristic, thus it is useful to consider other ways to understand this scaling.  A natural language to do so is in terms of potentials which are complex.   By construction, a complex potential describes the elastic motion of the particles with a loss of flux due to inelasticity.  This implies that the Hamiltonian for the system is  energy-dependent.   Both the non-Hermiticity and the energy dependence arise from an underlying Hermitian and energy-independent Hamiltonian through the elimination of degrees of freedom.  We note here that as a matter of principle it is always possible to describe 2-body scattering in such a language.  That is,  there exists a complex  local  but possibly energy-dependent potential which, when put into some suitable relativistic generalization of the Schr\"odinger equation for two body scattering, accurately reproduces all two-body observables including the $S$-matrix.  Here, we argue that a complex potential treatment naturally gives rise to the scaling in   Eq.~(\ref{logSmat}).  The argument has two parts: i) that {\it if} the imaginary part of the complex potential is of order $N_c^1$ then the imaginary part of $\log(S)$ is also of order $N_c^1$ as given in Eq.~(\ref{logSmat}) and  ii) that the imaginary part of the complex potential  is, in fact,  of order $N_c^1$.  

Let us begin by focusing on the first part of the argument. It starts with the fact that there exists a complex potential which accurately reproduces all two-body observables.  Let us assume that in this system, the mass is of order $N_c$, the system is Witten kinematics and the real and imaginary part of the complex potentials are of order $N_c$.  For simplicity, let us assume that nucleon spin and flavor play no role (an assumption we revisit in the next section).  In essence, this is the model of Subsection \ref{toy}, supplemented by an imaginary part of the potential to account for meson emission.  

As in Subsection \ref{toy}, the large $N_c$ limit automatically pushes the system into the semi-classical regime.  It is worth discussing precisely what this means in the context of a complex potential.   Again one can associate the phase with Hamilton's characteristic function $w$ for an equivalent classical problem.    However, when using a complex potential, $w$ is in general complex.  The imaginary part of $w$---the imaginary part of the phase---characterizes the  inelasticity.   As in the case of a real potential, the dominant classical trajectory associated with the quantum wave function is one for which the quantum mechanical phase is stationary leading to constructive interference.  This is given by the solution of the Hamilton-Jacobi equations using the real part of $w$.  The values for the real and imaginary part of  $w$ for the trajectory which solves these with fixed $l$ and $p$ in Witten kinematics gives the real and imaginary part of the phase shift for large $N_c$ dynamics.  If both the real and imaginary parts of the potential are of order $N_c$ it is clear that so is $w$ and one obtains the scaling exactly as in  Eq.~(\ref{logSmat}).

The second part of the argument is that imaginary part of the complex potential is of order $N_c^1$.    A useful way to understand this is via the formalism of Feshbach projection operators \cite{Feshbach}.  In principle one can always reproduce the results of QCD  with a hadronic model.   We will take the Hamiltonian for this model to be Hermitian and independent of energy.     
The Hilbert space of such model in the two baryon sector can be divided into two pieces: those components containing no mesons and those containing at least one meson.  A states purely in one of  these sectors are denoted $|\psi\rangle_{\rm nm}$, and $|\psi\rangle_{\rm m}$, respectively.  By construction ${}_{\rm nm}\langle \psi|\psi' \rangle_{\rm m}=0$ and the most general state in the space is of the form $|\psi\rangle = \alpha |\psi\rangle_{\rm nm} + \beta |\psi\rangle_{\rm m} $ with $\alpha^2 + \beta^2 =1$.  The Feshbach projection operator $P$ projects onto the no meson space, while $Q = 1-P$ projects onto the space containing at least one meson: $P |\psi\rangle = \alpha |\psi\rangle_{\rm nm}$ , $Q|\psi\rangle = \beta |\psi\rangle_{\rm m} $.  It is easy to show formally that if $\psi$ solves a time dependent Schr\"odinger equation $H |\psi \rangle = E |\psi \rangle$, then $|\psi\rangle_{\rm nm}$ satisfies:
\begin{equation}
\begin{split}
\left ( H_{\rm in} +V_{\rm out}(E) \right )|\psi\rangle_{\rm nm} & = E|\psi\rangle_{\rm nm} \; \; \; {\rm with}\\
 H_{\rm in}=2 M+ T+V_{\rm in} &\equiv P H P \,\,,\\
 V_{\rm out}(E) &\equiv  PHQ \frac{1}{E-H}QHP  \; .
\label{Feshbach1}\end{split}\end{equation}
 $H_{\rm in}$ is the Hamiltonian acting directly in the no-meson space.  It consists of a mass term for the two baryons, a  kinetic energy and a potential associated with interactions which do not push the system out of the no-meson space.  We will assume  consistently with standard counting that $V_{\rm in}$ is local, with a strength of order $N_c$ and range independent of $N_c$---or at any rate can be cast into that form by a unitary transformation acting in the two-nucleon space.
 
 $V_{\rm out}(E)$ corresponds to an interaction taking the system into the space containing mesons, propagating in that space and then another interaction taking the system back into no meson space; it is explicitly energy dependent.  Unfortunately, as written  $V_{\rm out}$ is ill-defined since the inverse operator depends on boundary conditions.  The correct choice of boundary conditions should build in the fact that we are interested in  propagation forward in time.  This is achieved by adding an infinitesimal imaginary part to the denominator of the second term:
\begin{equation}
 V_{\rm out}(E)=PHQ  \frac{1}{E-H+ i \epsilon}QHP  
\; .
\label{Feshbach2}\end{equation}

Note that from its structure $V_{\rm out}(E)$ satisfies a dispersion type relation {\it at the operator level} in the two-nucleon space.  In particular 
\begin{equation}
{\rm Re} \left( V_{\rm out}(E) \right) = \int d E' \,\, {\cal P} \left(\frac{{\rm Im} \left( V_{\rm out}(E) \right) }{\pi (E-E')} \right ) \,\,,
\end{equation}
where $ {\cal P}$ indicates principal part.   
This dispersion structure means that {\it if} ${\rm Im} \left( V_{\rm out}(E) \right) $ is a local (or nearly local) operator in $x$ for the two-nucleon space with a range  independent of $N_c$ and strength proportional to $N_c$---or can cast into such a form via a unitary transformation acting in the space containing at least one meson---then so is  ${\rm Re} \left( V_{\rm out}(E) \right) $.  Since we expect the real part to satisfy such a behavior to reproduce the Witten scaling behavior (which as note in earlier does apply to energy dependent interactions), it is highly plausible that the imaginary part does so as well.  But this was precisely the last condition needed to show that imaganiary part of the phase-shift, ({\it i.e.} the inelastic part of the S matrix) is consistent with Eq.~(\ref{logSmat}).

\subsubsection{Potential as a meson exchange}

A general argument was just provided as to why it is natural for the imaginary part of the potential (and hence the imaginary part of the phase shift) in Witten kinematics to scale as $N_c^1$.  However, it is useful to see how this comes about in a concrete model.  Here  we  use  a simple meson-exchange picture that respects the underlying large $N_c$ dynamics of QCD to infer the $N_c$ scaling of both the real and the imaginary part the potential.

We start by reviewing some basic principles of a meson-exchange picture from the point of view of the $N_c$ scaling 
\cite{KaplanSavage, KaplanManohar} and then study the role of the imaginary part.   For simplicity, we will be focusing on the case of nonrelativistic kinematics.  Various cancellations which occur in a transparent way in nonrelativistic systems are obscure in relativistic ones.  Note, however, that at large $N_c$, the nucleon mass mass is of order $N_c$.   Thus systems exist in which the incident nucleon velocity $v \ll 1$ but for which $M_N v^2 \sim N_c  \Lambda_{\rm QCD}^2$ and hence which are of order $N_c$ above the elastic threshold and can emit an order of $N_c$ mesons during a scattering process.  Thus, this restriction should not alter the basic $N_c$ counting.  Similarly,
for simplicity, we again will do the detailed analysis assuming  only scalar mesons and only one species of these.  We will also consider only the simplest type of Yukawa interaction.   This simplification will not alter the $N_c$ counting and the algebra is simple and traceable.  Had we included mesons with other quantum numbers, the emergent spin-flavor contracted symmetry 
\cite{GervaisSakita84-1, GervaisSakita84-2, DashenManohar93-1, DashenManohar93-2, DashenJenkinsManohar94, DashenJenkinsManohar95} would be needed to ensure the sorts of cancellations which naturally emerge with scalars \cite{BanerjeeCohen...}.    

\begin{figure}
\begin{center}

\includegraphics[width=1.7in]{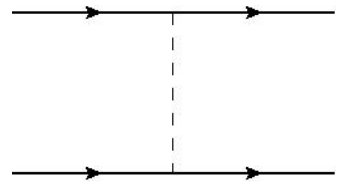}

\caption{A Feynman diagram of the one meson exchange part of the potetial.}
\label{1piex}

\end{center}
\end{figure}

The Feynman diagram representing one-meson exchange is in the Fig. \ref{1piex}. The $N_c$ scaling of this diagram is given
by the $N_c$ scaling of the coupling constant standing in the meson-nucleon-nucleon vertex, which is of order $g \sim N_c^{1/2}$ 
\cite{WittenNC}.
Note  that the characteristic momentum flowing through the meson is of order $N_c^0$ since the meson mass is independent of $N_c$ which means that the propagator will cut off momenta which are characteristically  larger.    Since there are two vertices in 
Fig. \ref{1piex}, the amplitude coming from this diagram---which can be identified with the one-boson exchange contribution to the potential---is of order $N_c$.  
Note that this is precisely the scaling a nucleon-nucleon potential should obey according to the standard counting rules.
Note that with elastic kinematics where the energy  transfer is zero the kinematics are such that there is no intermediate state of a two nucleons plus meson and hence the imaginary part of the potential is zero.

\begin{figure}
\begin{center}

\includegraphics[width=2.5in]{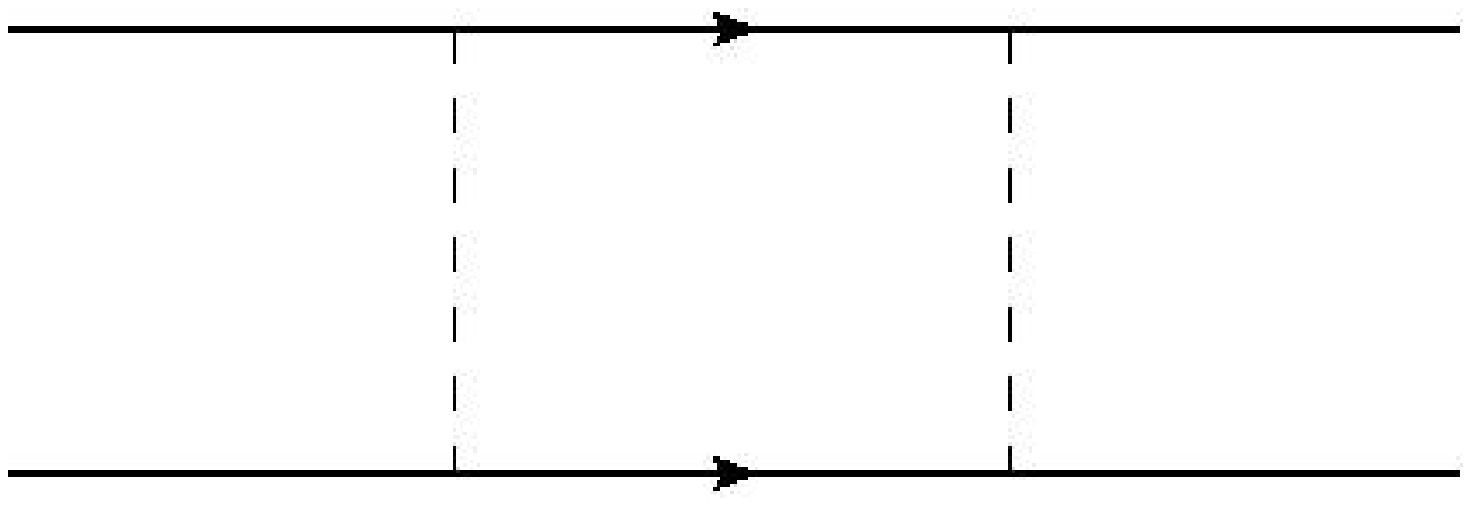}
\vspace{0.2in}

\includegraphics[width=2.5in]{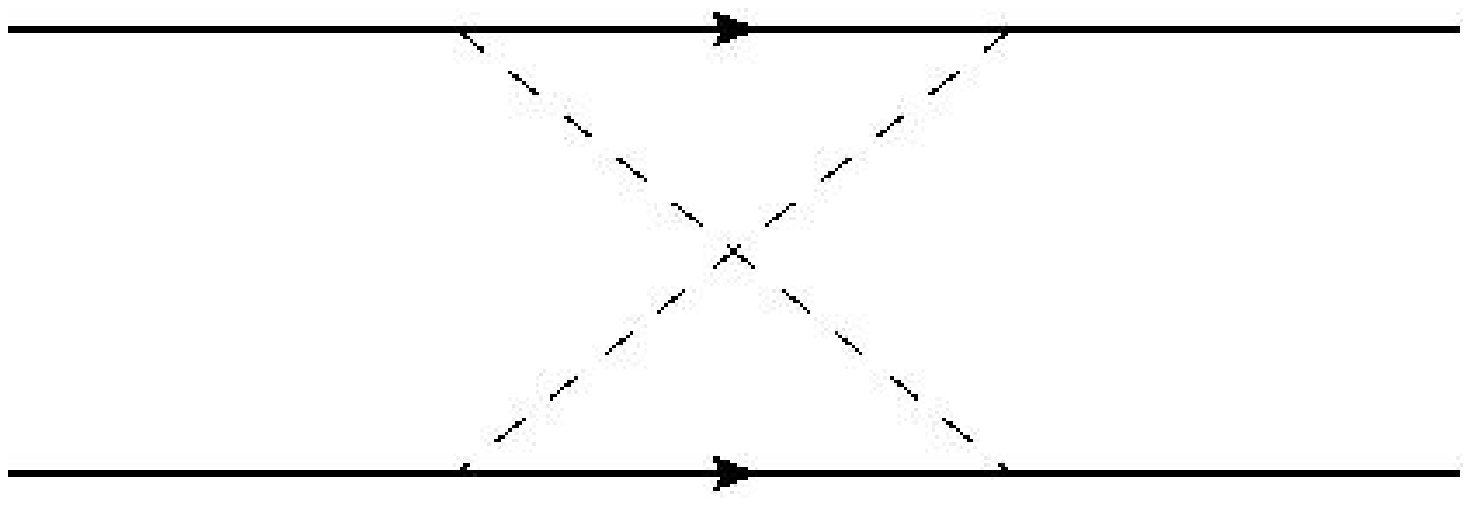}

\caption{A Feynman diagrams of the two-meson exchange interaction, one boxed and one crossed.}
\label{2piex}

\end{center}
\end{figure}

On the other hand, the situation gets more complicated in a case of two meson exchange, as was discussed
extensively in  a paper of Banerjee et. al  \cite{BanerjeeCohen...}.
Relevant diagrams are summarized in Fig. \ref{2piex}.
The naive counting of powers of $N_c$ leads to an overall scaling $N_c^2$, which is a problem, since the potential
should be of order $N_c^1$. However, a deeper analysis resolves the apparent problem in two steps.
First, one observes that the baryon-pole contribution to the amplitude of the box diagram is exactly of the
same form as a first iterate of a Lippmann-Schwinger equation with a potential given by one meson exchange.
As such, it does not contribute to the potential so that double counting is avoided.
Since potential is the quantity that should be of order $N_c$, the first iterate containing two potentials
is then of order $N_c^2$, as it should be.
In the second step one is able to show that the meson-pole contribution of the box diagram (the retardation effect)
and the crossed diagram differ only by sign at the leading order in $1/N_c$ expansion. Thus,
they subtract exactly and the remaining piece, which is not an iterate and therefore genuinely enters the potential,
does not spoil the overall $N_c$ behavior.
This exact cancellation at  leading order occurs only to the extent that $v \ll 1$ and for this reason we focus on nonrelativistic dynamics.
The connection between the potential picture and a meson-exchange picture is rather subtle  at the level of
three meson exchange as discussed in  \cite{Cohen3MesEx1}. But with a sensible definition, the potential can be made to scale in a consistent way  \cite{Cohen3MesEx2}.

The imaginary part of the potential emerges naturally in a meson exchange picture.  For simplicity we will focus on two-meson exchange 
since it is the lowest order interaction in which an imaginary part emerges.  We will also not focus on the box and crossed box diagrams since these are subtle in that there are superleading contributions of order $N_c^2$ and the leading order result comes from a cancellation.  
It was discussed in the preceding section that a consistent meson-exchange picture for a nucleon-nucleon potential can be constructed. However, one can also construct a two-meson exchange diagram with a non-zero imaginary part in nonrelativistic kinematics and which does not depend on any cancellations.

\begin{figure}
\begin{center}

\includegraphics[width=1.7in]{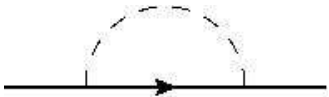}

\caption{A Feynman diagram of the one meson loop correction to the nucleon propagator.}
\label{barpropag}

\end{center}
\end{figure}

Before investigating how an imaginary part emerges in the potential, let us  first look at the simplest possible one loop diagram: the one
meson loop correction to a baryon propagator, which is shown in Fig. \ref{barpropag}.

In the nonrelativistic regime, the loop integral reads
\begin{eqnarray}
I(k^0-M)&=&\int \frac{{\rm d}^4 l}{(2\pi)^4} \frac{i}{l^2 -m_{\pi}^2 + i \epsilon}\, \frac{i}{k^0-M+l^0+ i \epsilon} \nonumber\\
&=& {\rm Re}(I) \, +  \, i\, \pi \, \theta(k^0-M-m_{\pi}) \,, \label{selfenergy}
\end{eqnarray}
where $k^0$ is the zeroth component of the nucleon four-momentum.
As one sees directly from (\ref{selfenergy}), the integral acquires the imaginary part when the energy is sufficient
to excite a meson. It is in agreement with the naive picture of what an imaginary part of an amplitude means: that a new particle can be produced.

\begin{figure}
\begin{center}

\includegraphics[width=2.5in]{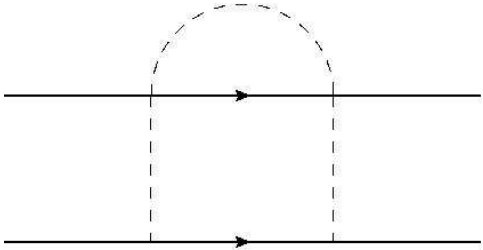}

\caption{A Feynman diagram which contributes to the imaginary part of the potential in the leading order.}
\label{impartseagull}

\end{center}
\end{figure}

It is natural to consider two-meson-exchange diagrams which contain this self energy as a subdiagram.  Many of these require complicated cancellations of the sort discussed above.  However, there exists a simple case where no cancellations occur.     It is a two-meson exchange interaction with one extra loop on one of the nucleons.  This is shown in Fig. \ref{impartseagull}. 



 It is easy to see that both the real and imaginary parts scale as $N_c^1$.  The key point is that the self energy subdiagram is connected to the system via a 2-nucleon-2-meson vertex.  By standard Witten counting this scales as $N_c^0$.  Again, the typical momentum transfer through exchanged mesons is order $N_c^0$.  However, there is also a possible energy transfer of the same order, yielding an on-shell configuration of two nucleons and a meson.  Thus both the real and the imaginary parts are of order $N_c^1$ as one expects for a potential yielding an S-matrix consistent with the standard Witten counting.

There are other  possible two-meson exchange diagrams which contain imaginary parts.  In many of these the nominal $N_c$ counting is higher than $N_c^1$.  However, we have verified that all of these contain cancellations analogous to that seen between the retardation effect in the box and crossed diagrams.  Thus, we expect that in this kinematic regime the leading order will be of order $N_c$ for the imaginary part.
This model result is suggestive rather than decisive.   Apart from the simplifications noted at the outset---that the analysis is in a particular kinematic regime and of a particularly simple model---the argument here was only for one type of diagram.  Nevertheless it is useful check that it is consistent with our expectation from more general arguments.

\subsubsection{Potential at a quark and gluon level}

The previous  discussion was formulated in the language of mesons and baryons. It was argued somewhat heuristically,
that the proposed $N_c$ scaling of both real and imaginary part of the potential, i.e. that they are both proportional to $N_c$, holds
in the nucleon-nucleon scattering. In this section, we argue that the previous result is natural  also at the level of quarks and gluons.  We will again focus on diagrams which at the hadronic level look like Fig.~\ref {impartseagull}.

\begin{figure}
\begin{center}

\includegraphics[width=2.5in]{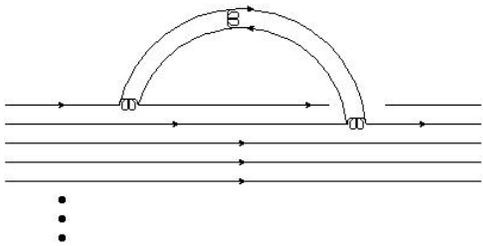}

\caption{A leading order diagram for the nucleon self energy.}
\label{NNmeson}

\end{center}
\end{figure}

\begin{figure}
\begin{center}

\includegraphics[width=1.3in]{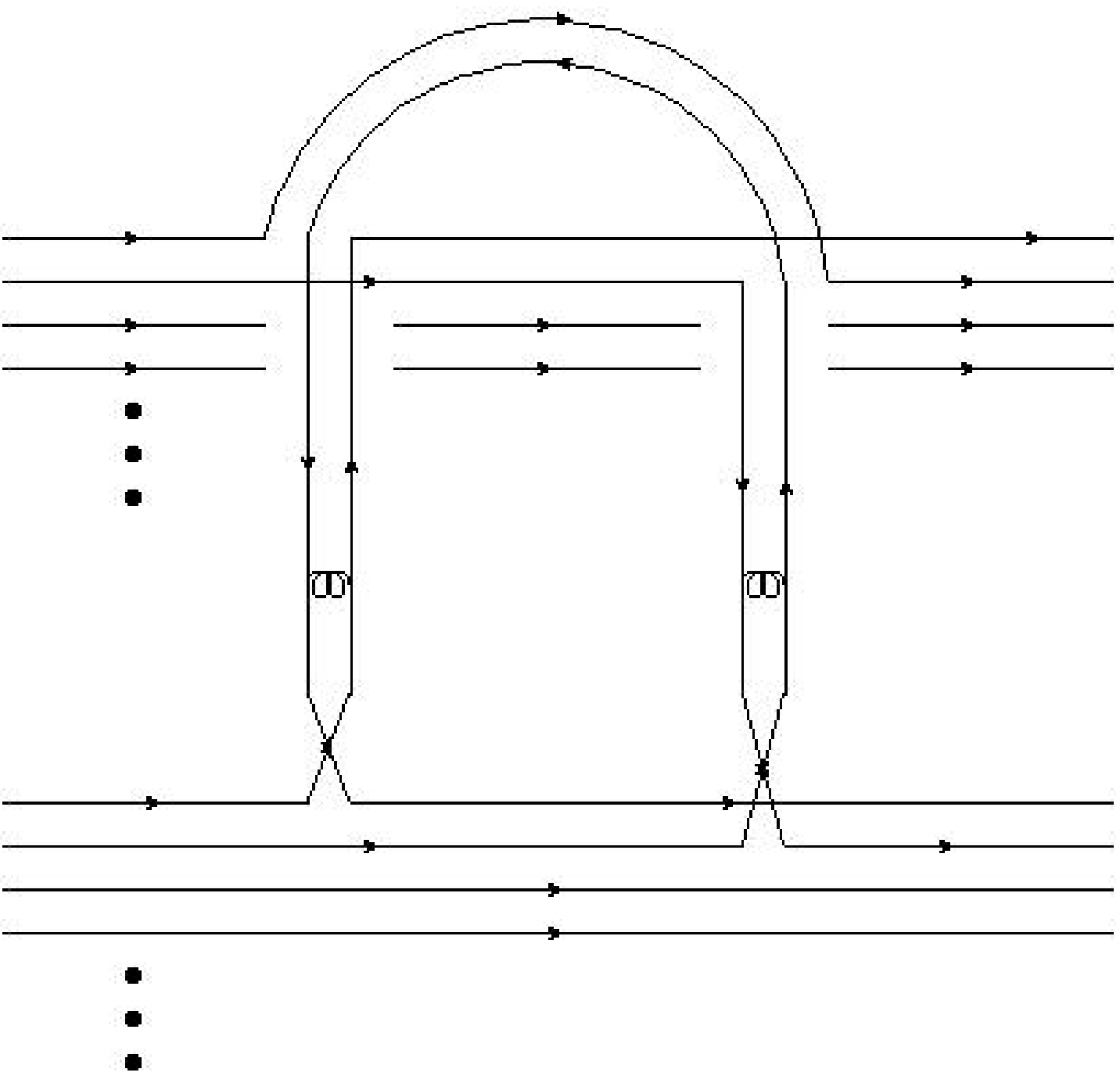}
\hspace{0.1in}
\includegraphics[width=1.3in]{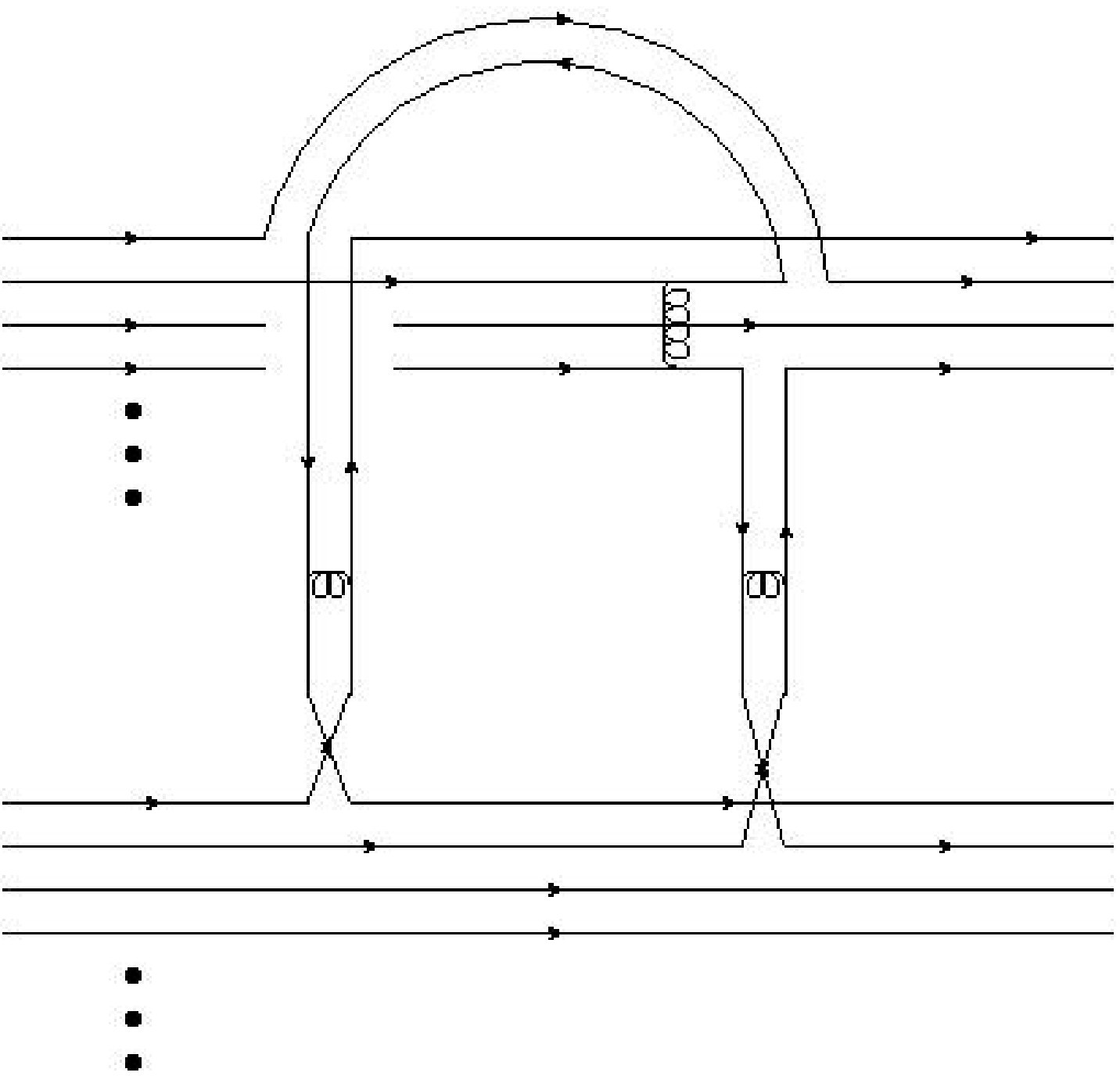}

\caption{Diagrams at the quark-gluon level analogous to Fig.~\ref{impartseagull}.}
\label{analog}

\end{center}
\end{figure}

Let us start by looking at the self-energy diagrams containing meson loops such as in Fig.~\ref{NNmeson}.  Note that this diagram was drawn to emphasize the fact that Feynman propagators contain backward going components, so that even without quark loops meson loops contribute at order $N_c^1$, the leading order in the $1/N_c$ expansion \cite{CohenLeinweber}.  With the appropriate kinematics such a meson can go on-shell.  Now let us turn to Fig.~\ref{analog}, which shows diagrams analogous to Fig.~\ref {impartseagull}.  Note that the quark-antiquark pairs playing the role of the exchanged meson, couple to the same quark line in the upper nucleon or to quarks connected to it via gluons.  The reason that this is required is the following:  following the first exchange the nucleons can propagator either in their ground state or in an excited state.  If the first  exchange leaves both nucleon in their ground state, then it acts like an iterate of the potential.  Thus such contributions to two meson exchange   cannot be part of the potential itself.  However, if one of the nucleons is excited by pushing one quark into an excited state that quark must be deexicted in the second exchange.  This happens in the second exchange.   It is a simple exercise in $N_c$ counting to verify that both of these diagrams are of order $N_c$ when combinatorics is appropriately taken into account.  The key point is that the restriction of the exchanges to the same quark line in the upper nucleon or to quarks connected to it via a gluon reduce the $N_c$ counting by a factor of $1/N_c$ compared to a naive counting.  In the case where the exchanges couple to the same quark line, there is a reduction of a combinatoric factor of $N_c$.  In the case where quarks are connected via a gluon there is no reduction in combinatoric factors but there is an extra factor of $g^2 \sim 1/N_c$.

\section{Spin and Flavor}

Up to this point the analysis conveniently skipped the fact that nucleons have spin and isospin degrees of freedom.  This enabled us to formulate the problem of finding the S-matrix as one of finding the real and imaginary parts of phase shifts for given orbital momentum partial wave.  A variety of arguments  indicate that these scale as $N_c^1$.  When spin and flavor are put back into the problem the S-matrix is a matrix for each angular momentum and flavor channel.  This  could represent a major complication since the analysis presented above was semi-classical in nature while spin and flavor degrees of freedom are intrinsically quantum mechanical in nature.  One might worry that the gist of the argument could be spoiled for this intrinsically quantum system.    

There is an additional complication.  At large $N_c$, there is an emergent $SU(2N_f)$ spin-flavor symmetry \cite{GervaisSakita84-1, GervaisSakita84-2, DashenManohar93-1, DashenManohar93-2, DashenJenkinsManohar94, DashenJenkinsManohar95} and the nucleon is part of a tower of nearly degenerate stable baryon states.  If, for simplicity, one restricts attention to a world with 2 flavors, this tower consists of states with $I=J$ with $J \sim N_c^0$  (the case of three or more flavors adds technical complications but will not change the fundamental result).  
The Skyrmion \cite{Skyrme1} is a concrete realization of this in the context of a simple model.  The Skymion naturally gives isolated baryons in terms of hedgehog configurations which correlate directions in space and isospace.   Since these break both rotational and isorotational symmetry, (iso)rotated skymions are also solutions to the equations and are equally good baryons. Such rotations depend on three Euler angles.   At a quantum level, these hedgehogs are interpreted as superpositions of low lying states in the band with $I=J$.  Note that the classical description in terms of Euler angles emerges naturally at large $N_c$ regardless of whether one is using a Skyrme model---it holds directly in QCD.  There is a strong analogy to deformed states in mean-field descriptions of nuclei.  As with deformed nuclei one can project on to states with good quantum numbers to extract the physically relevant states.  The process is in analogy with Pieirls-Yaccoz projection \cite{PieirlsYaccoz}  and, as in that case, one does this by integrating the collective degrees of freedom---the Euler angles---weighted by Wigner D-matrices and an appropriate normalization constant.

Fortunately, this second complication allows one to deal with the first.  The emergence of collective degrees of freedom associated with spin and flavor means that one can deal with these degrees of freedom in an essentially classical way.  The approach is similar to the one in Ref.~\cite{CohenGelman02}.  At large $N_c$ one can consider the initial states in the scattering are not baryons with $I $ and $J$ quantum numbers but rotated hedgehogs specified by Euler angles.  This is allowable in a quantum sense in that, as in all physical scattering processes, we consider incident wave packets coming in from asymptotic distances; in this case the wave packets contain a superposition of incident baryon states.  Now, if we consider taking the large $N_c$ prior to the long-distance limit in setting up the scattering problem the hedgehog remains coherent as it enters the scattering region and a classical picture emerges.  The scattering process takes time of order $N_c^0$.  As the two hedgehogs leave the interaction region (we are now doing the analysis for the case in which no mesons have been emitted)  they will presumably have been rotated but the amount of the rotation is not expected to scale with $N_c$ but to go as $N_c^0$.  The reason for this is that the torques exerted on one hedgehog by the other should be of order $N_c^1$ by standard Witten counting but so is the moment of inertia and the time the torque is exerted is of order $N_c^0$.  Of course ultimately the wave packets associated with rotated hedgehogs will disperse and the various baryon types separate but if one takes the large $N_c$ limit at the outset the time scale for this to happen goes to infinity.  Thus the rotation angles experienced by the hedgehogs becomes arbitrarily well defined in the large $N_c$ limit and are given by a classical analysis.  The phase associated with this classical motion in both real space and collective coordinate space---including the imaginary part---is again given by Hamilton's principle function.  The arguments are essentially the same as   in the case of a spinless particles.

We will denote this phase as $\delta_l(A^i_1,A^i_2;E)$ where $A^i$ denotes the initial set of collective angles for each particle where $A$ has the norm $\int d A =(2\pi^2)^{-1}$ .  We note that we do not need to specify the final collective angles since the classical nature of the dynamics means that these are fixed by the initial ones.   Thus there are functions $A_1^f(A^i_1,A^i_2)$ and $A_2^f(A^i_1,A^i_2)$ which specify outgoing collective variables.  Note that from the arguments given above we expect  $\delta_l(A^i_1,A^i_2:E)$ to be of order $N_c$ in Witten kinematics with $l=N_c \tilde{l} $ and $E=N_c \tilde{E} $.  Thus we write $\delta_l(A^i_1,A^i_2;E) =N_c \tilde{\delta}_{\tilde{l}}(A^i_1,A^i_2;\tilde{E})$

Finally, we need to extract the nucleon-nucleon S-matrix from the classical result from these by projection.  These are given by 
\begin{widetext}
\begin{equation}
S=\int dA_1 dA_2 \,  D^{1/2}_{m_1,{m^I_1}(A_1)} \, D^{1/2}_{m_2,{m^I_2}(A_2)} \, D^{*1/2}_{m_1,{m^I_1}(A_1^f(A^i_1,A^i_2))} \, D^{*1/2}_{m_2,{m^I_2}(A_2)} \, e^{i N_c \tilde{\delta}_{\tilde{l}}(A^i_1,A^i_2;\tilde{E})}
\end{equation}
\end{widetext}
Analytically continuing $\delta$ into the generalized complex plane allows this integral to be evaluated by steepest descents and this yields matrix elements which are power law in $N_c$ times $\exp( N_c \tilde{\delta}_{\tilde{l}}(B^i_1,B^i_2;\tilde{E})$ where $B_1$,$B_2$ are the complex values for $A_1$, $A_2$ where the derivative of $\delta$ vanishes.  This result yields Eq.~(\ref{logSmat}) including the full  spin and isospin dependence.
  
\section{Conclusion}

This paper has given a variety of arguments in support of Eq.~(\ref{logSmat})---the notion that both the real and the imaginary parts of the logarithm of the S-matrix scale with $N_c$.  While all of these arguments were to some degree heuristic, they all give the same result.  Together, they comprise a rather compelling argument for the validity of Eq.~(\ref{logSmat}).  These arguments all exploit the semi-classical nature of large $N_c$ dynamics in Witten kinematics.  While the simplest formulation of these arguments was given in simplified models in which spin and flavor were neglected, it was shown in the last section that the collective spin and flavor rotations at large $N_c$ allows these degrees of freedom to be included in the semi-classical analysis.  

\bigskip
\emph{Acknowledgments} \\
This work was supported by the U.S.~Department of Energy
through grant DE-FG02-93ER-40762. \\
V.K. also acknowledge the support of JSA/Jefferson Lab Graduate Fellowship.

\end{document}